\documentclass[%
reprint,
 amsmath,assume,
 aps,
]{revtex4-2}

\usepackage{graphicx}
\usepackage{dcolumn}
\usepackage{bm}

\newcommand{\new}[1]{{{#1}}}

\begin{document}


\title{Lightwave-controlled relativistic plasma mirrors}

\author{Marie Ouillé$^{1,2,3}$}
\author{Jaismeen Kaur$^{1,3}$}\email{jaismeen.kaur@ensta-paris.fr}
\author{Zhao Cheng$^{1}$}
\author{Stefan Haessler$^{1}$}\email{stefan.haessler@cnrs.fr}
\author{Rodrigo Lopez-Martens$^{1}$}\email{rodrigo.lopez-martens@ensta-paris.fr}

\affiliation{$^{1}$Laboratoire d'Optique Appliqu\'ee, Institut Polytechnique de Paris, ENSTA-Paris, Ecole Polytechnique, CNRS, 91120 Palaiseau, France,}
\affiliation{$^{2}$Ardop Engineering, 6 All. Annie Fratellini, 33140 Villenave-d'Ornon, France}
\affiliation{$^{3}$These authors contributed equally to this work.}

\date{\today}

\begin{abstract}
	We report on attosecond-scale control of high-harmonic and fast electron emission from plasma mirrors driven by relativistic-intensity near-single-cycle lightwaves at kHz repetition rate. By controlling the waveform of the intense light transient, we reproducibly form a sub-cycle temporal intensity gate at the plasma mirror surface, leading to the observation of extreme ultraviolet spectral continua, characteristic of isolated attosecond pulse generation. We also observe the correlated emission of a waveform-dependent relativistic electron beam, paving the way from towards fully lightwave-controlled dynamics of relativistic plasma mirrors.
\end{abstract}

	\maketitle
	
	The generation of attosecond light pulses has enabled the real-time observation of the fastest electronic processes in matter. Although new generation schemes have emerged \cite{Hassan:16, Franz2024}, the currently preferred route to produce attosecond pulses remains High-Harmonic Generation (HHG) from the gas phase \cite{Takahashi:13, Nayak:18}. The mechanism is now well understood, the technology is mature (as indicated by the emergence of turn-key, tabletop HHG beamlines), and it has been successfully used in several pioneering applications unveiling previously unobservable attosecond-scale dynamics \cite{Borrego-Varillas:2022}.
	However, it is intrinsically limited to low intensity ($\sim 10^{15}\,\rm W/cm^2$) driving pulses and the maximum laser-to-XUV conversion efficiency is currently $\lesssim10^{-4}$~\cite{Takahashi:13, Nayak:18}. To circumvent those limitations, HHG from relativistic plasma mirrors has long been recognized as a promising route \cite{Heissler:12} owing to the predicted percent-level conversion efficiencies \cite{Mikhailova:12} (although the measured efficiencies are $\sim10^{-4}$ so far~\cite{Rodel:12, Jahn:19}), their compatibility with relativistic-intensity ($> 10^{18}\,\rm W/cm^2$) driving lasers, as well as the production of chirp-free attosecond pulses~\cite{Chopineau:21} with high spatial coherence~\cite{Dromey:09, Leblanc:17}. 
	
	The nonlinear interaction leading to the emission of XUV light being repeated once (in the case of surface harmonics) or twice (in the case of gas harmonics) per optical cycle, trains of attosecond pulses spaced by one or half a laser period, respectively, are usually generated. To gate the process and generate isolated attosecond pulses (IAP), different techniques based on largely available multi-cycle drivers have been explored, such as polarization gating \cite{Baeva:06, Yeung:15}, wavefront rotation \cite{Wheeler:12, Vincenti:12} and two-color gating \cite{Zhang:20, Smith:21, Blanco:17}. However, these methods are costly in terms of achievable focused laser intensity, and it is much more efficient to gate the process using single-cycle duration laser pulses with controlled carrier-envelope phase (CEP). Routinely used for IAP generation from the gas phase \cite{Baltuska:03} for well over a decade now, this gating technique has thus far only been applied to relativistic plasma mirrors through the binning and post-processing of data generated with randomly fluctuating laser waveforms \cite{kormin:2018, Jahn:19, Boehle:20}. 
	
	HHG from plasma mirrors is also known to be accompanied by the emission of electrons \cite{ChopineauPRX_couplings, Haessler:21}, injected at specific phases in the reflected laser pulse \cite{Thevenet:16}, in the form of attosecond bunches \cite{Zhou:21}. Typically, in the case of multi-cycle driving pulses, depending on their initial energy and time of injection during the pulse, these electrons can either be isotropically scattered due to the laser ponderomotive potential, leading to the observation of a doughnut-shaped electron distribution centered about the specular direction far away from the interaction point, or undergo direct vacuum laser acceleration (VLA) and further gain kinetic energy before being ejected along the polarization plane, leading to the observation of an electron density peak slightly offset from the specular direction \cite{VLA}. 
	
	In this letter, we present experimental results obtained on the correlated high-harmonic and electron emission from relativistic plasma mirrors driven by CEP-controlled sub-1.5-cycle laser pulses: for the first time, via CEP gating, we are able to reproducibly generate XUV continua and control relativistic electron beaming with attosecond precision, thereby paving the way towards the reliable generation of synchronized isolated attosecond light and electron bunches from relativistic lightwave-driven plasma mirrors.
	
	Experiments were conducted using the \textit{Salle Noire} laser driver~\cite{SN2Laser}, developed in-house at \new{\textit{LOA}} 
	, which routinely delivers multi-mJ energy sub-4 fs pulses (FWHM) at a central wavelength $\lambda_0 \approx 780$ nm, corresponding to less than 1.5 optical cycles, with a temporal contrast ratio $> 10^{10}$ up to 15 ps and better than $10^7$ up to 3 ps before the pulse peak. A shot-to-shot CEP stability of $\approx$ 380 mrad rms can be maintained for \new{$\sim$ hour durations \cite{Huijts:22}. The CEP is measured and locked relative to an unknown offset value at the 1-kHz laser repetition rate by a \textit{Fringeezz} device with feedback to a \textit{Dazzler} (both Fastlite). It can be varied while keeping all other laser parameters identical.}
	
	\begin{figure}[tb]
		\centering
		\includegraphics[width=\columnwidth]{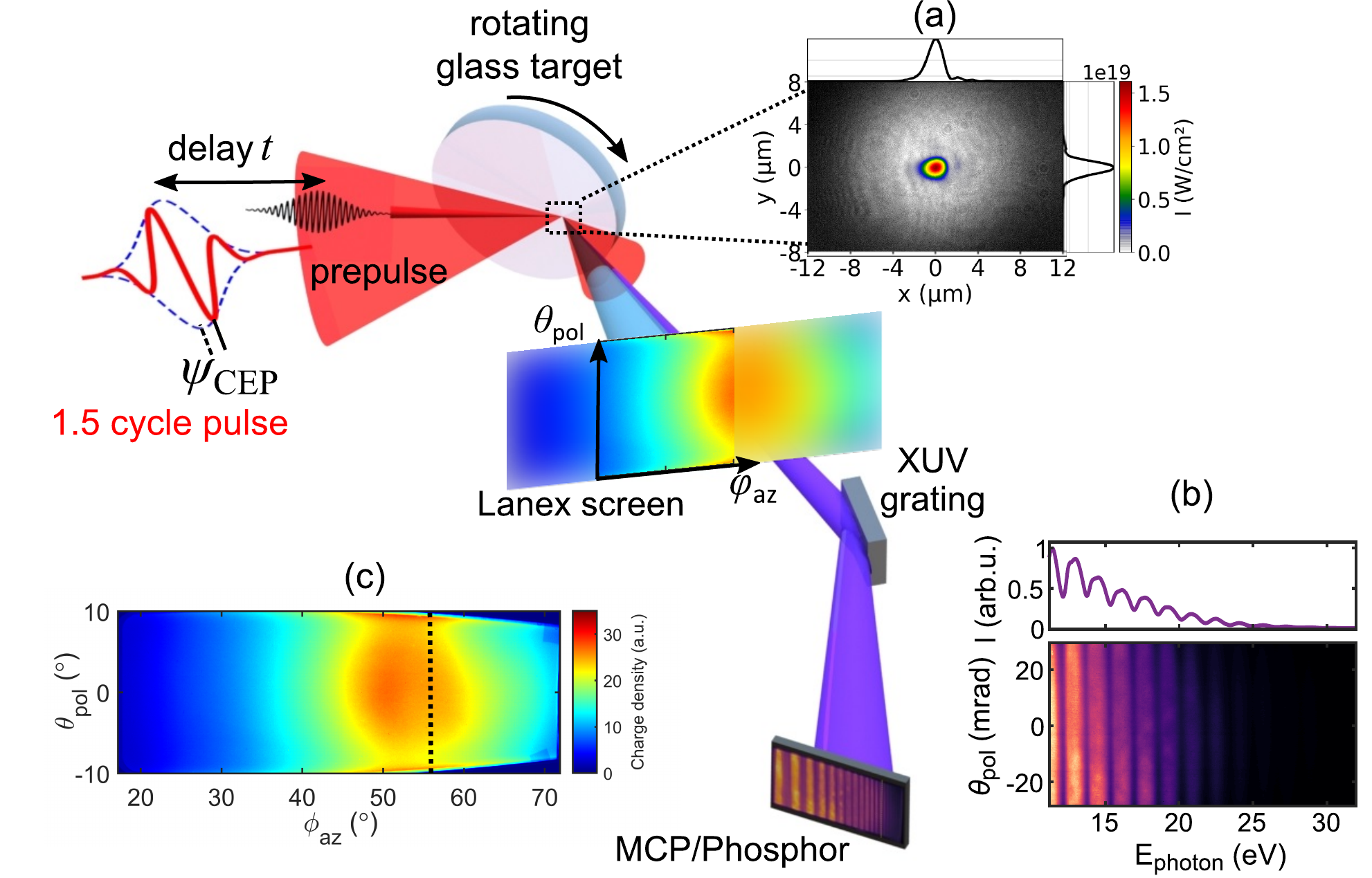} 
		\caption{Schematic drawing of the experimental setup. Insert~(a) shows the experimentally measured focused beam profiles on target for both the Prepulse (gray-color scale) and Main-pulse (multi-color scale). Insert~(b) shows a typical angle-resolved XUV spectrogram (top: vertically integrated spectrum). Insert~(c) shows a typical electron angular charge distribution recorded with the wide-angle Lanex screen. The dashed-black line indicates the specular direction.}
		\label{fig:exp_setup}
	\end{figure}
	
	The experimental setup \cite{Kaur:23} is shown in fig.~\ref{fig:exp_setup}. The laser pulses are made to interact with plasma mirrors generated at the surface of rotating solid fused silica targets \cite{Borot:14}. We use a time-delayed Prepulse, obtained by picking-off a small fraction ($\approx$ 4 $\%$ of the energy) of the Main-pulse, in order to initiate and control plasma expansion at the target surface. Both p-polarized pulses are tightly focused on target at an incidence angle of $55^\circ$ by a f/1.3, $30^\circ$ off-axis parabola. As shown in the fig.~\ref{fig:exp_setup}(a), the Prepulse and Main-pulse are focused down to $\approx10\:\mu$m and $\approx1.5 \times 1.7\:\mu$m FWHM spot size, respectively. Considering the Main-pulse energy on target to be approximately 2.5 mJ and assuming there are no spatio-temporal couplings, this corresponds to a relativistic peak intensity of $I_\mathrm{peak} = 1.8 \times 10^{19}\,\rm W/cm^2$ \textit{i.e.} a normalized vector potential $a_0 = \lambda_0 [\mu \rm{m}] {\sqrt{ I_{\rm{peak}} \rm{[W\,cm^{-2}]}   /1.37 \times 10^{18} }} \approx 2.8$. 
	
	The spatio-spectral XUV content of the reflected beam (photon energies ranging from 10 eV to 40 eV) is measured in the specular direction ($\phi_{az}=55^\circ$ with respect to the target-normal) with a flat-field concave XUV grating and an MCP-phosphor screen assembly imaged on a CCD camera (fig.~\ref{fig:exp_setup}(b)). A scintillating screen (Lanex) covered with a 13-\textmu m aluminum foil and imaged onto another CCD camera allows simultaneous detection of electrons with kinetic energy $> 150\:$keV in the range, $30^\circ < \phi_{az} < 48^\circ$ with respect to the target normal. A 0.5-mm pinhole and a pair of permanent magnets can be inserted in front of the Lanex screen in order to measure the energy spectra of electrons emitted at $\phi_{az} \approx 38^\circ$ and $\theta_{pol} \approx 0^\circ$, which allows simultaneous XUV detection. Alternatively, a larger Lanex screen can be inserted to image the full electron angular charge distribution covering $18^\circ < \phi_{az} < 72^\circ$ with respect to the target normal (fig.~\ref{fig:exp_setup}(c)), which now blocks XUV detection and prevents any direct correlation measurement with XUV generation. Nevertheless, different experimental reports confirm that electron emission and HHG are correlated in the relativistic regime \cite{ChopineauPRX_couplings, Haessler:21, Kaur:23}. No electron spectra could be recorded in this configuration due to the size of the Lanex assembly.  
	
	\begin{figure}[t!]
		\centering
		\includegraphics[width=0.85\columnwidth]{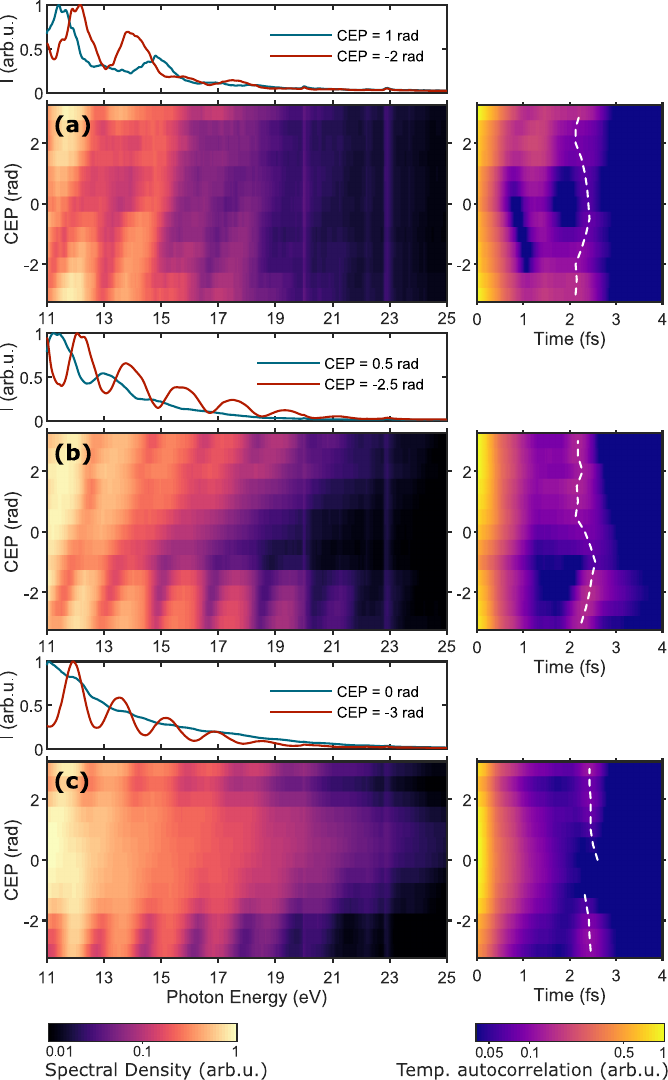} 
		\caption{Vertically-integrated HHG spectra as a function of relative CEP for Prepulse lead times of: -1 ps (a), 1 ps (b), and 2 ps (c). Above each frame are two line-out spectra measured for opposite laser waveforms ($\pi$ CEP offset). Each acquisition averaged over 100 consecutive laser shots (100~ms). To the right of each frame, we show the temporal auto-correlation of the HHG emission in the shown spectral range. The white dashed line marks the position of the satellite peak near 1~laser period \new{where such a peak could be determined}. The scans were made in succession while the CEP-locking loop remained active. \new{Equal} relative CEPs thus correspond to the same driving waveforms in each panel.}
		\label{fig:data1}
	\end{figure}
	
	\begin{figure*}[tb]
		\centering
		\includegraphics[width=0.95\textwidth]{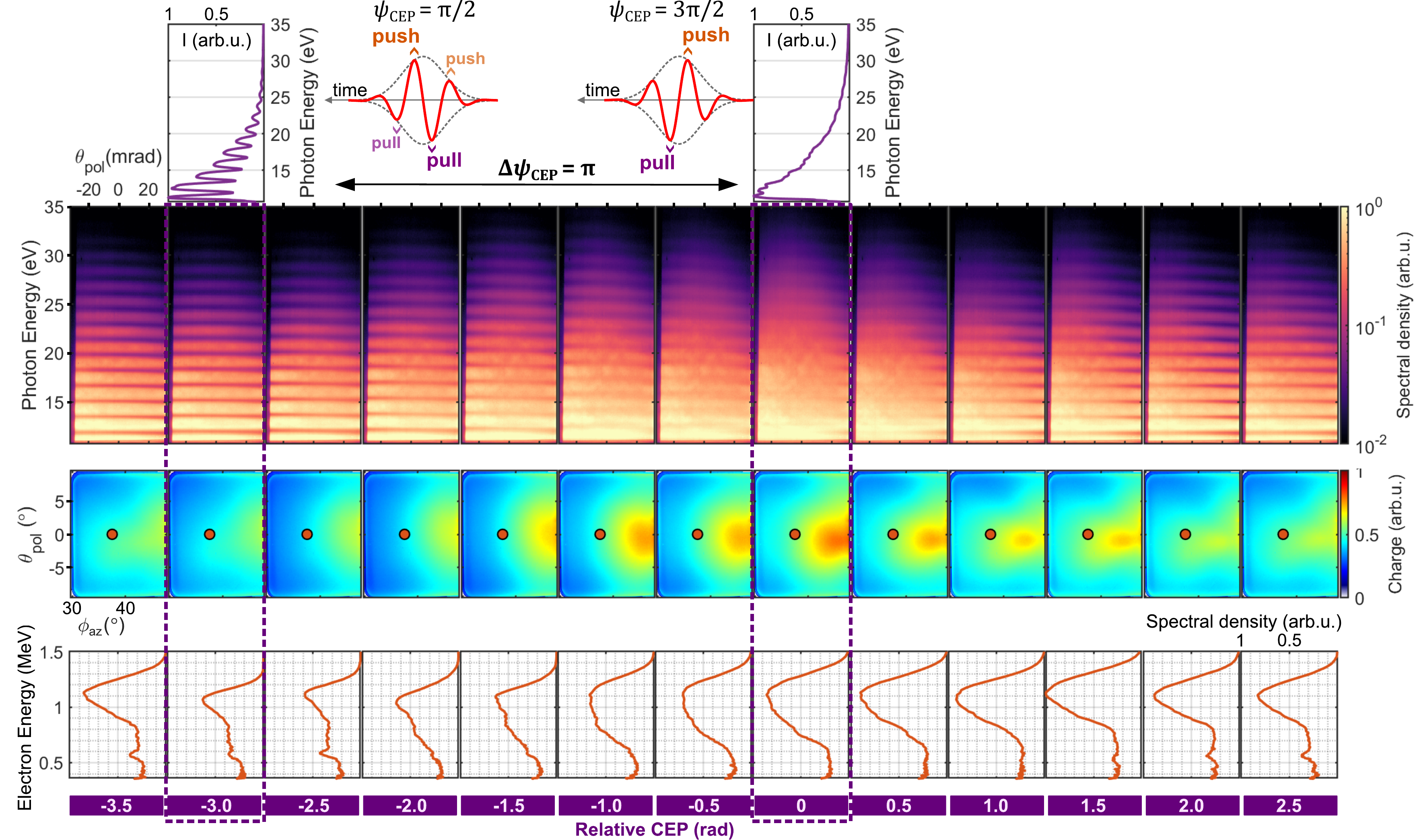}
		\caption{Simultaneous acquisition of XUV spectra (top panel), electron beam profiles (middle panel), and electron spectra (bottom panel) for different relative CEP values. Each acquisition averaged over 100 consecutive laser shots (100~ms). The red dot in the middle panel indicates the point of measurement of the electron spectrum. The dashed boxes indicate the two opposite waveforms in the electron push-pull scenario: at the top, the angle-integrated HHG spectrum is shown for the two opposite driving waveforms and corresponding push-pull scenarios at absolute CEP values, $\psi_{\textrm{CEP}} = \pi/2$ and $3\pi/2$. The formation of an XUV continuum can clearly be resolved in the case of a single push-pull sequence for a given relative CEP (step size $< 0.5\:$rad).}
		\label{fig:data2}
	\end{figure*}
	
	In order to estimate the surface plasma gradient scale length $L_g = L_0 + c_s \tau$, where $L_0 \approx 0.01 \lambda_0$ is the estimated contribution of the Main-pulse finite temporal contrast, and $\tau$ the delay between the Prepulse and the Main-pulse, we experimentally measured the plasma expansion speed $c_s$ using spatial domain interferometry \cite{SDI, Kaur:23} and found $c_s = 22\ $nm/ps. In fig.~\ref{fig:data1}, we show the vertically-integrated XUV spectra recorded as a function of relative CEP for three different values of $\tau$: -1~ps, 1~ps, and 2~ps, corresponding to respectively $L_g \approx L_0$, $0.04\lambda_0$ and $0.07\lambda_0$. We observe clear CEP and $L_g$-dependent spectral structures of the harmonics. Assuming a nearly flat spectral phase~\cite{Chopineau:21}, these can be linked to the periodicity of the corresponding attosecond temporal profiles, apparent in the temporal auto-correlations obtained as the modulus of the Fourier transform of each HHG spectrum in the shown spectral range and also shown in fig.~\ref{fig:data1}. For $L_g \approx L_0$ (fig.~\ref{fig:data1}a), the slower spectral modulation on top of the harmonic orders probably results from the presence of both coherent wake emission~\cite{Quere_cwe:06} and relativistic HHG. Their different emission times~\cite{Chopineau:21} then create 2 attosecond pulses per $\approx2.5$-fs laser period, which appears as an additional peak near 1.3~fs in the auto-correlations. As the plasma gradient softens to $L_g\approx0.04\lambda_0$ (fig.~\ref{fig:data1}b), relativistic HHG becomes dominant, and only the satellite peak near the $\approx2.5$-fs laser period remains in the auto-correlation. Its CEP-dependent shift over $\approx0.3\:$fs likely results from a CEP-dependent plasma denting evolution over the pulse duration~\cite{Jahn:19}. Finally, in panel (c), there is interestingly no clear temporal shift of the satellite peak (i.e., the harmonic modulation period is nearly constant).
	
	Finally, we find that a 2-ps prepulse delay ($L_g\approx 0.07\lambda_0$, Fig.~\ref{fig:data1}c) maximizes the spectral extent (for $\tau > 3\:$ps, both harmonic yield and spectral extent drop), which is close to the optimal gradient of $L_g\approx0.1\lambda_0$ typically observed for relativistic HHG \cite{Kahaly:13,Haessler:21}. Only for this optimum gradient do we observe clear XUV spectral continua for a given relative CEP ($0\:$rad in Fig.\ref{fig:data1}c), characteristic of IAP generation. Clearly modulated spectra are observed when the relative CEP is offset by $\pi$. Interestingly, the corresponding $\approx2.5$-fs satellite peak in the temporal auto-correlations shifts by no more than $\approx0.1\:$fs before falling below the noise level.
	
	We repeated these measurements at the optimal density gradient ($L_g \approx 0.07\lambda_0$), but this time simultaneously recording both the XUV spectrum and the electron angular charge distribution as a function of relative CEP (fig. \ref{fig:data2}). The locked CEP was varied randomly to eliminate any systematic experimental bias. Here again, an XUV spectral continuum is clearly produced for a certain relative CEP value ($0~$rad), whereas the most modulated spectrum appears upon adding a $\approx \pi$ phase offset ($-3~$rad). This $\pi$ periodicity is reminiscent of our previous observations via CEP-tagging and is very well reproduced by numerical simulations predicting compression of the lightwave into a sub-0.4 fs IAP, potentially containing a third of the incident pulse energy \cite{Boehle:20}, without any spectral filtering required. Physically, this CEP-dependence can be understood through a three-step process \cite{Thevenet:16, Boehle:20}: (1) push, (2) pull, and (3) emission of XUV light, which occurs only once in the optimal-gating-case of $\psi_{\textrm{CEP}}=3\pi/2$ and twice in the case of $\psi_{\textrm{CEP}}=\pi/2$ (see fig.\ref{fig:data2}). We also clearly observe, for the first time, a closely correlated waveform-dependence of the electron emission: the highest electron charge is obtained in correlation with the XUV continuum in which case a clear charge density peak appears close to the specular direction, while the lowest charge is produced by the opposite laser waveform ($\pi$ CEP offset), in which case no density peak is observed. 
	
	\begin{figure}[tb]
		\centering
		\includegraphics[width=1\columnwidth]{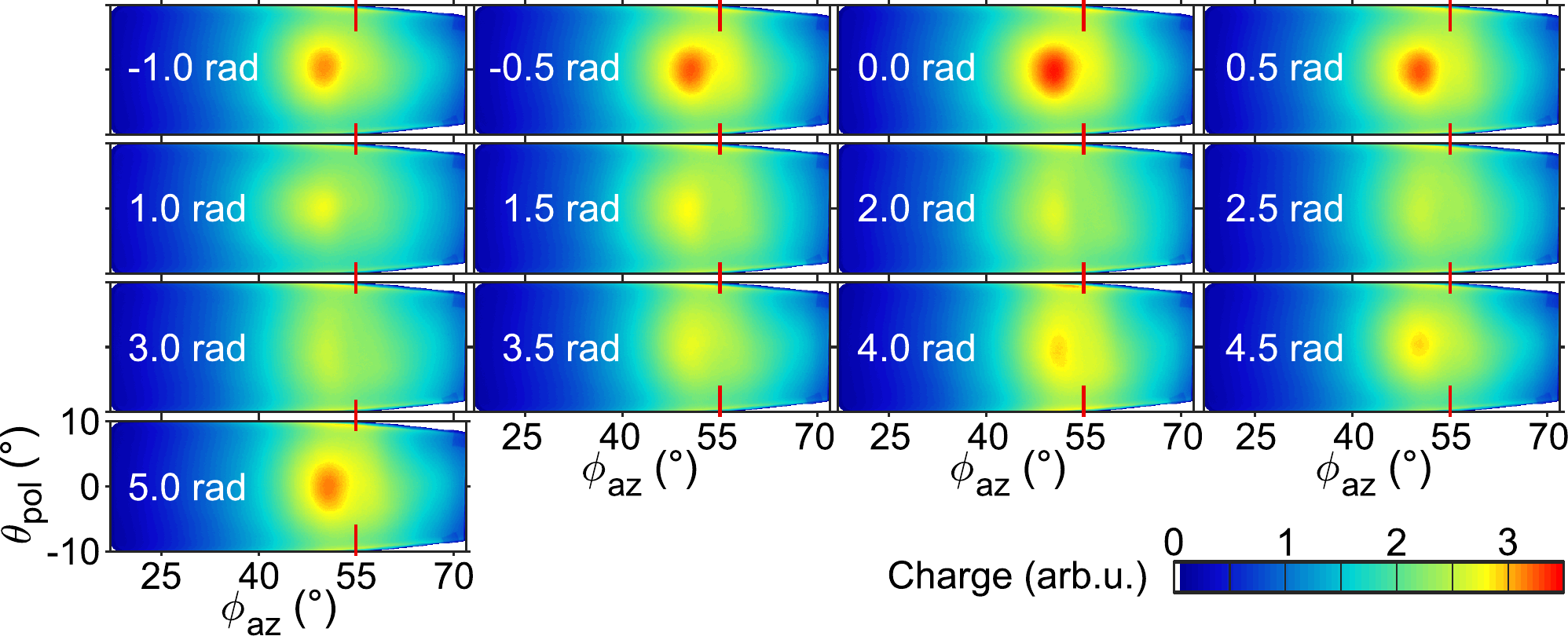}
		\caption{Angular electron charge distribution for the locked CEP values indicated in the panels, measured at $\tau = 2\:$ps Prepulse lead time. Each image averaged over 100 consecutive laser shots (100~ms). \new{Red ticks indicate the specular direction.}}
		\label{fig:electronBeamProfile}
	\end{figure}

	Measuring the electron energy spectrum (point of measurement indicated by the red dot in the middle panel of fig.~\ref{fig:data2}) confirms the clear influence of the CEP on the electron acceleration mechanism. The dashed boxes in fig.~\ref{fig:data2} highlight the acquired electron spectra under similar conditions for the two opposite laser waveforms: In the right box, for a given relative CEP ($0\:$rad), we are in the scenario in which the push-pull mechanism only occurs once, as indicated by the production of an XUV continuum. Interestingly, this corresponds to the production of an electron energy spectrum with a single broad peak at $\approx$ 1 MeV. As the CEP is shifted by $\pi$ (left box, CEP$=-3~$rad), a more modulated electron spectrum is observed, with the appearance of a shoulder at lower energies. These features could be the indication of the phase-locked injection and subsequent VLA of either one (in the optimal-gating case of $\psi_{\textrm{CEP}}=3\pi/2$) or two (in the case of $\psi_{\textrm{CEP}}=\pi/2$) attosecond relativistic electron bunches. \new{In the latter case, one bunch undergoes less efficient VLA than the other.} These preliminary results should be confirmed in a more detailed experimental study. Unfortunately, our setup only allowed us to measure the spectrum at the electron beam periphery, indicated by the red dot in the middle panel in Fig.~\ref{fig:data2}. We expect higher energies in the beam center, in particular in the optimal-gating scenario (right box).
	Our observation is nonetheless the first experimental evidence of CEP-controlled emission of relativistic electrons from plasma mirrors. 
	
	Fig.~\ref{fig:electronBeamProfile} shows the full electron beam measured under similar conditions with the wide-angle Lanex screen, as the relative CEP is cycled through $2\pi$. The electron angular charge distribution exhibits the same CEP dependence as in the previous measurements (middle panel in fig.~\ref{fig:data2}), with maximum charge peak appearing at a relative CEP of $0~$rad and a minimally charged beam (approximately half the total integrated charge) when the CEP is shifted by $\pi$. \new{Although the distribution and direction of the electron beam only subtly changes, for a relative CEP of $0~$rad, the less divergent ``VLA beam'', centered on $\phi_\mathrm{az}\approx 50^\circ$, is most dominant over the more divergent contribution centered near the specular direction. This dominance gradually fades with CEP until for a relative CEP of 3~rad, a small sign of ponderomotive blowout appears around the specular direction. This would indicate that when $\psi_{\textrm{CEP}}=\pi/2$, one of the electron bunches is not injected with the adequate phase-energy combination for efficient VLA, leading to a lower final energy and some ponderomotive scattering.} We plan to measure the spatially-resolved electron spectrum across the full beam in order to visualize CEP-dependent fine structure that could provide more insight into the attosecond-scale electron acceleration dynamics.        
	
	In conclusion, we report the first experimental evidence of the CEP dependence and control of correlated high-harmonic and relativistic electron emission from plasma mirrors driven at kHz repetition rate by controlled near-single-cycle lightwaves. In particular, we can produce XUV spectral continua on demand by forming a singular sub-cycle temporal intensity gate at the plasma mirror surface, which in light of previously reported experimental observations of the very high spatio-temporal quality of the emitted attosecond pulses~\cite{Chopineau:21,Dromey:09,Leblanc:17} represents a major step towards the realization of isolated XUV attosecond pulses with record-intensity at kHz repetition rate, which we aim at verifying in the future.
	
	Data underlying the results presented in this paper are available in Ref.~\cite{Kaur:24_CEPdataset}.
	
	\begin{acknowledgments}
	This work was supported by the Agence Nationale pour la Recherche (ANR-11-EQPX-005 ATTOLAB, ANR-14-CE32-0011-03 APERO, ANR-22-CE30-0005-01 BANDITO); European Research Council (ERC FEMTOELEC 306708, ERC ExCoMet 694596); LASERLAB-EUROPE (H2020-EU.1.4.1.2. grant agreement ID 654148), R\'egion Ile-de-France (SESAME 2012-ATTOLITE).
	\end{acknowledgments}

\bibliography{refs}
\end{document}